\documentclass[%
 reprint,
showpacs,preprintnumbers,
nofootinbib,
 amsmath,amssymb,
 aps,
 pra,
longbibliography,
floatfix,
noeprint
]{revtex4-1}

\pdfoutput=1

\usepackage{graphicx}% Include figure files
\usepackage{dcolumn}% Align table columns on decimal point
\usepackage{bm}% bold math
\usepackage{dsfont}
\usepackage[dvipsnames]{xcolor}
\usepackage{hyperref}% add hypertext capabilities
\hypersetup{
  colorlinks   = true, %Colours links instead of ugly boxes
  urlcolor     = MidnightBlue, %Colour for external hyperlinks
  linkcolor    = blue, %Colour of internal links
  citecolor   = blue %Colour of citations
}

\def\d{\partial}
\def\+{\dagger}
\def\la{\langle}
\def\ra{\rangle}

\def\id{\mathds{1}}

\begin{document}

%\preprint{APS/123-QED}

\title{Symmetry-protected coherent relaxation of open quantum systems}% Force line breaks with \\

\author{Moos van Caspel$^1$}
  \email{M.T.vanCaspel@uva.nl\\}
\author{Vladimir Gritsev$^{1,2}$}%

\affiliation{%
 $^1$Institute of Physics and Delta Institute for Theoretical Physics, University of Amsterdam\\ Science Park 904, 1098 XH Amsterdam, The Netherlands\\
 $^2$Russian Quantum Center, Skolkovo, Moscow 143025, Russia
}%

\date{\today}% It is always \today, today,

\begin{abstract}
We compute the effect of Markovian bulk dephasing noise on the staggered magnetization of the \mbox{spin-$\frac{1}{2}$} XXZ Heisenberg chain, as the system evolves after a N\'eel quench. For sufficiently weak system-bath coupling, the unitary dynamics are found to be preserved up to a single exponential damping factor. This is a consequence of the interplay between $\mathbb{PT}$ symmetry and weak symmetries, which strengthens previous predictions for $\mathbb{PT}$-symmetric Liouvillian dynamics. Requirements are a non-degenerate $\mathbb{PT}$-symmetric generator of time evolution $\hat{\mathcal{L}}$, a weak parity symmetry and an observable that is anti-symmetric under this parity transformation. The spectrum of $\hat{\mathcal{L}}$ then splits up into symmetry sectors, yielding the same decay rate for all modes that contribute to the observable's time evolution. This phenomenon may be realized in trapped ion experiments and has possible implications for the control of decoherence in out-of-equilibrium many-body systems.

% \begin{description}
% %Secondary publications and information retrieval purposes.
% \item[PACS numbers]
% %May be entered using the \verb+\pacs{#1}+ command.
% %You may use the \texttt{description} environment to structure your abstract;
% %use the optional argument of the \verb+\item+ command to give the category of each item.
% \end{description}
\end{abstract}

\maketitle

%\tableofcontents

%Ideas for introduction:
%- Many-body dissipative systems: why?
%- Applications: quantum computing?
%- Most research focused on long-time behavior
%- Short-time behavior only possible in specific cases (quadratic, fermionic)
%- Important to study spectrum in context of observables
%- Symmetries play an important role, determining what part of the spectrum contributes to what observable
%- Specific interplay of symmetries leads to coherent relaxation of observable

\section{Introduction}\label{section:introduction}

The theory of open quantum systems has a long history, finding countless applications in quantum optics, nanotechnology, quantum information and other fields of physics \cite{Breuer}. Particularly in the past decade, there has been a drive to apply this formalism to the realm of many-body quantum systems. Rapid developments in the fields of cold atoms and quantum computation form major incentives to improve our understanding of the interaction between a quantum system and its environment. Whether one is interested in shielding a system from decoherence or driving it towards a specific steady state, the theoretical challenges are largely the same and they are formidable.

Analytical methods to tackle dissipative many-body systems are few and far between. Most efforts are focused on Markovian baths, allowing a formulation in terms of a Lindblad master equation. Exact solutions have been given for quadratic fermionic systems \cite{Eisler2011,Prosen2010,Prosen2010a,Torres2014,Znidaric2010}, but this excludes most bulk dissipation in spin systems. Various algebraic methods have been used to solve Lindblad equations \cite{Ringel2012,Bolanos2015,Mesterhazy2017}, requiring the unitary and dissipative parts to form a closed algebra. Finally, very specific models have been mapped to integrable closed systems, solvable by Bethe ansatz \cite{Essler_prosen}. On the other hand, numerical approaches are typically restricted to either very short time scales or to the infinite-time limit. Properties of the full relaxation process are surprisingly difficult to probe, but the presence of symmetries can simplify the problem.

Symmetry structures in the context of open quantum systems have been studied mostly in relation to stationary states and conserved quantities  \cite{AlbertLiang,Nicacio2016,Wilming2017,Manzano2014}. They are closely tied to the theory of decoherence-free subspaces and subsystems, which are considered promising candidates for quantum memory \cite{Zanardi1997}. However, symmetries can also play an important role in the dynamics at shorter time scales. These dynamics satisfy the Lindblad equation and are generated by the Liouvillian superoperator $\hat{\mathcal{L}}$, acting on the space of linear operators. Symmetries allow for a separation of operator sectors, which split up the spectrum of $\hat{\mathcal{L}}$. For an observable, this means that large parts of the spectrum may not contribute toward the time evolution of its expectation value, depending on the symmetry properties of the observable. We use this phenomenon, combined with the spectral structure of a $\mathbb{PT}$-symmetric Liouvillian \cite{Prosen2012}, to study a scenario where dissipation affects the dynamics of observables in a predictable and coherent manner.

In general, adding a non-unitary part to a system's time evolution introduces many new time scales, corresponding to the decay rates of the different modes of the time evolution's generator $\hat{\mathcal{L}}$. For a generic system, all of these modes will contribute at intermediate times, affecting the dynamics in a highly nontrivial way. In the system we study --- a \mbox{spin-$\frac{1}{2}$} XXZ Heisenberg chain affected by bulk dephasing noise --- some observables are protected by symmetry from all but one of the system's decay rates. The result is an overall damping factor such that the unitary dynamics are preserved for weak system-bath coupling. This surprising effect should be experimentally measurable and may be relevant for the control of decoherence in many-body quantum gates.

The structure of the paper is as follows: in section \ref{section:lindblad}, we review the Lindblad master equation and its spectral properties. Section \ref{section:symmetries} details the different types of symmetries and their interplay in Liouvillian dynamics. Section \ref{section:xxz} describes how these symmetries apply to the \mbox{spin-$\frac{1}{2}$} XXZ Heisenberg chain with bulk dephasing. Finally, we study the staggered magnetization after a N\'eel quench in section \ref{section:neel}, as an example of symmetry-protected coherent relaxation. The appendix shows a detailed perturbation theory calculation of corrections to the staggered magnetization.

\section{Lindbladian time evolution}\label{section:lindblad}
Markovian dynamics can always be described in terms of a Lindblad master equation of the form
\begin{align}
  \begin{split}
  \frac{\d \rho}{\d t} &= -i \left[ H, \rho \right]\\ &+ \gamma \sum_i \left( L_i \rho L^\+_i - \frac{1}{2} L^\+_i L_i \rho - \frac{1}{2} \rho L^\+_i L_i  \right),
\end{split}
\end{align}
where $\gamma>0$ is the system-bath coupling strength and $L_i$ are the so-called jump operators, which encode the interaction between the system and the bath. This form can typically be derived from a microscopic theory by integrating out the bath and applying several approximations, such as the Born-Markov and the Rotating Wave approximation \cite{Breuer}. It is often convenient to write the Lindblad equation in superoperator form:
\begin{align}
  \frac{\d \rho}{\d t} = \hat{\mathcal{L}}\rho ~~~ \Rightarrow ~~~ \rho(t) = e^{t \hat{\mathcal{L}}} \rho(0).
\end{align}
where the \emph{Liouvillian superoperator} $\hat{\mathcal{L}}$ is a trace, hermiticity and positivity-preserving linear map, such that it maps one density matrix to another. Superoperators act on the space $\mathcal{B}(\mathcal{H})$, consisting of all linear operators acting on the Hilbert space of quantum states $\mathcal{H}$. In turn, $\mathcal{B}(\mathcal{H})$ itself can be treated as a Hilbert space with the Hilbert-Schmidt inner product: $(A,B) \equiv \operatorname{tr}(A^\+ B)$. In what follows, we will be particularly concerned with non-degenerate Liouvillians, which are diagonalizable\footnote{This is not generally true for Liouvillians. However, a non-diagonalizable superoperator can be expressed in a comparable form using a Jordan decomposition. Most of the following qualitative statements will still hold true in this situation, although one can get power-law contributions to the expansion of $\rho(t)$. See e.g.\ \cite{Prosen2010,Pletyukhov2010}} and can therefore be written as a spectral decomposition:
\begin{align}
  \begin{split}
  \hat{\mathcal{L}}\rho &= \sum_m \lambda_m \operatorname{tr}(v_m^\+ \rho)\, u_m\\  &\Rightarrow ~~~ \rho(t) = \sum_m e^{t \lambda_m} \operatorname{tr}(v_m^\+ \rho(0))\, u_m,
  \end{split}
\end{align}
where $\hat{\mathcal{L}} u_m = \lambda_m u_m$ and $\hat{\mathcal{L}}^\+ v_m = \lambda_m^* v_m$ such that $\operatorname{tr}(v_m^\+ u_m) = 1$. Since $\hat{\mathcal{L}}$ is not Hermitian, its left and right eigenmodes are not equal and its eigenvalues $\lambda_m$ are generally complex. Furthermore, $\operatorname{Re}(\lambda_m) \leq 0$ or else $\rho(t)$ would blow up in the infinite time limit. Thanks to Brouwer's fixed point theorem, there must be at least one zero eigenvalue $\lambda_0 = 0$. The corresponding eigenmode $u_0$ is known as a \emph{steady state} of the time evolution. Symmetries can result in multiple steady states, as we will show in section \ref{section:symmetries}. One can also have persistent oscillations with $\lambda \neq 0$ on the imaginary axis, but these are rare and will not be discussed further in this work. All other modes $u_m$ with $\operatorname{Re}(\lambda_m) \neq 0$ are known as \emph{decay modes}, and they must be traceless operators.

Studying the spectrum of the Liouvillian can tell you much about the non-unitary dynamics. One particular quantity of interest is the \emph{dissipative gap}, defined as $\Gamma = \operatorname{min}_{\text{decay modes}}\{|\operatorname{Re}(\lambda_m)|\}$. The gap determines the longest timescale in the system. At long times, generic observables decay exponentially at rate $\Gamma$. Expanding the time evolution of the expectation values of observables yields:
\begin{align}
  \la O(t) \ra = \operatorname{tr}(O \rho(t)) = \sum_m e^{t \lambda_m} \operatorname{tr}(v_m^\+ \rho(0)) \operatorname{tr}(O u_m). \label{eq:obs_spectral}
\end{align}
At sufficiently long times, the dissipative gap dominates all higher decay modes and determines the rate at which the steady state is approached. For some systems the gap may close in the thermodynamic limit, leading to algebraic decay \cite{CaiBarthel}. But as we will see, the presence of symmetries can throw a wrench into this simplified picture. Each symmetry sector has its own gap and the decay rates can be vastly different between observables.

Lastly, it is illuminating to consider the spectrum of the Liouvillian for a closed system, i.e.\ $\gamma = 0$. The eigenvalues are purely imaginary and given by $\lambda = i(\epsilon_i - \epsilon_j)$, corresponding to the eigenmodes $|\psi_i\ra \la\psi_j|$ with $H|\psi_i\ra = \epsilon_i |\psi_i\ra$. There is a degeneracy at $\lambda = 0$, the size of the Hilbert space, as projectors onto energy eigenstates are naturally stationary. If we then turn on a weak dissipation, degenerate perturbation theory shows that these diagonal modes $|\psi_i\ra \la\psi_i|$ will hybridize and their eigenvalues will spread out. In the case of Hermitian jump operators $L_i^\+ = L_i$ or in the presence of $\mathbb{PT}$ symmetry (see section \ref{section:symmetries}), they will stay on the real axis.

\section{Symmetries in Hilbert space, Liouville space and beyond}\label{section:symmetries}
In the context of unitary time evolution, discrete symmetries are relatively straightforward. They are typically generated by a Hermitian operator $O$, acting on the Hilbert space $\mathcal{H}$, such that $[H, O] = 0$. As a result, energy eigenstates are simultaneously eigenstates of $O$. The Hilbert space can therefore be separated into blocks, one for each eigenvalue of $O$, which are preserved under unitary time evolution. If there are multiple, mutually commuting symmetries, then there will be subblocks within each symmetry block.

When adding a dissipative, non-unitary part to the time evolution, this story becomes slightly more complicated \cite{AlbertLiang}. Symmetry on the level of the Hilbert space $\mathcal{H}$, as described above, still exists and we will call this a \emph{strong symmetry}, following Ref.\ [\onlinecite{Buca2012}]. In the case of Lindbladian evolution, the operator $O$ should not only commute with $H$, but also with each jump operator individually: $[L_i,O] = 0 ~ \forall i$. Once again the Hilbert space separates into blocks. Of course non-unitarity will produce mixed states, but it only mixes states within the same symmetry block.

This block structure of $\mathcal{H}$ can be lifted to the space of linear operators $\mathcal{B}(\mathcal{H})$, which we will call Liouville space. To make this more precise, consider $n$ symmetry blocks $\mathcal{U}_i$ that form a partition of $\mathcal{H}$. We can then partition $\mathcal{B}(\mathcal{H})$ into $n^2$ blocks $\hat{\mathcal{U}}_{i,j}$ spanned by operators of the form $|\psi\ra \la \phi|$ with the states $|\psi\ra \in \mathcal{U}_i$ and  $|\phi\ra \in \mathcal{U}_j$. Because of the strong symmetry, the $n$ `diagonal' blocks\footnote{These should \emph{not} be thought of in the sense of a block-diagonal matrix. For example, one can block-diagonalize the Liouvillian superoperator $\hat{\mathcal{L}}$, in which case \emph{all} blocks $\hat{\mathcal{U}}_{i,j}$ will be on the diagonal. Instead, these `diagonal' blocks relate to the diagonal matrix elements of operators.} $\hat{\mathcal{U}}_{i,i}$ must each have their own steady state. In rare cases, some `off-diagonal' blocks may also contain fixed points of the Liouvillian, yielding what is known as a decoherence-free subspace \cite{Lidar2003}.

However, one can have a block structure in Liouville space without the strict conditions of a strong symmetry. The only requirement for such a structure is a unitary superoperator that commutes with the Liouvillian: $[\hat{\mathcal{L}}, \hat{O}] = 0$, where unitarity is defined as preserving the Hilbert-Schmidt inner product. This is known as a \emph{weak symmetry} (or a dynamical symmetry in some literature \cite{Baumgartner2008}). Note that this requirement is immediately satisfied in case of a strong symmetry by defining $\hat{O}\rho = O \rho O^\+$. But a weak symmetry by itself does not imply the presence of multiple steady states. In general, only one block will contain the steady state, while the others are spanned by traceless decay modes. In section \ref{section:xxz}, examples of both weak and strong symmetries will be discussed in detail.

It is necessary to understand the symmetry structures of Liouville space when studying the time evolution of observables. Each of the Liouvillian's decay modes is confined to one symmetry block. If an observable has no components in a given symmetry block, then it is clear from Eq.\ (\ref{eq:obs_spectral}) that any decay modes in this block will not contribute towards the observable's time evolution. This can severely impact which parts of the spectrum are relevant, depending on the observables of interest. An extreme example is the staggered magnetization in the XXZ chain with dephasing, as we will see in section \ref{section:neel}. One more symmetry is needed to produce such a case, of a special type that acts on the Liouvillian superoperator itself. Table \ref{tab:symmetries} shows an overview of the three different types of symmetries.

$\mathbb{PT}$ symmetry in Lindbladian time evolution was first described in Ref.\ [\onlinecite{Prosen2012}]. Since it features prominently in the rest of this work, we will summarize its properties here but refer to the original paper for details. A Liouvillian is $\mathbb{PT}$-symmetric when it satisfies the condition:
\begin{align}
  &\hat{\mathcal{P}}\hat{\mathcal{L}}'\hat{\mathcal{P}} = - (\hat{\mathcal{L}}')^\+  \label{eq:ptsymmetry} \\
  &\hat{\mathcal{L}}' = \hat{\mathcal{L}} + \gamma \delta \hat{\id},~~~~~ \delta \equiv -\frac{\operatorname{tr} \hat{\mathcal{L}}}{\gamma \operatorname{tr}\hat{\id}} \label{eq:tracelessL}
\end{align}
where $\mathcal{L}'$ is the traceless part of the Liouvillian, $\hat{\mathcal{P}}$ is some (unitary) parity superoperator (with $\hat{\mathcal{P}}^2 = \id$) and the Hermitian adjoint is again defined using the Hilbert-Schmidt inner product. Since the unitary part of $\hat{\mathcal{L}}$ is traceless, $\operatorname{tr} \hat{\mathcal{L}}$ is proportional to $\gamma$, such that the scaling factor $\delta$ is dimensionless and does not depend on the coupling strength. In words, this is an  antisymmetry relating the adjoint of the traceless part of the Liouvillian to a parity transformation of the same. While this seems highly specific and not very physical, it can be considered a generalization of $\mathcal{PT}$-symmetric quantum mechanics \cite{Bender}. $\mathbb{PT}$-symmetric Liouvillians have some very nice properties and turn out to be surprisingly prevalent in spin systems \cite{Prosen2012a}.

\bgroup
\def\arraystretch{2}
\setlength\tabcolsep{5pt}
\begin{table}[t]
\begin{tabular}{l|c|c|c}
  \centering
   \emph{Symmetry~} & \emph{Acts on} & \emph{Condition} & \emph{Ex.\ XXZ} \\ \hline
   Strong & $\mathcal{H}$ & $ [H, O] = [L_i, O]$ = 0 & $\sum_i S^z_i$ \\
   Weak & $\mathcal{B}(\mathcal{H})$ & $[\hat{\mathcal{L}}, \hat{O}] = 0$ & $\hat{R}, \hat{F}$ \\
   $\mathbb{PT}$ & $\mathcal{B}(\mathcal{B}(\mathcal{H}))$ & $\hat{\mathcal{P}}\hat{\mathcal{L}}'\hat{\mathcal{P}} = - (\hat{\mathcal{L}}')^\+$ & $\hat{\mathcal{P}} \rho = F \rho$
\end{tabular}
\caption{Overview of different types of symmetries, acting on the hierarchy of Hilbert spaces. The last column shows the examples from the XXZ Heisenberg spin chain, as described in section \ref{section:xxz}. $\mathcal{B}(A)$ refers to the vector space of linear operators acting on space A.}
\label{tab:symmetries}
\end{table}
\egroup

The spectrum of a $\mathbb{PT}$-symmetric Liouvillian shows a second reflection symmetry axis in the complex plane, at $\operatorname{Re} \lambda = -\gamma \delta$. This is in addition to the reflection symmetry across the real axis, which is guaranteed by hermiticity conservation. In the absence of degeneracies and for sufficiently weak system-bath coupling, \emph{all} eigenvalues lie on these two axes. This can be seen by applying perturbation theory to the $\gamma = 0$ case, as mentioned at the end of section \ref{section:lindblad}. $\mathbb{PT}$ symmetry guarantees that the diagonal operators (in the energy eigenbasis) stay on the real axis when the dissipation is turned on \cite{Prosen2012}. Meanwhile, the off-diagonal coherences are confined to move along $\operatorname{Re} \lambda = -\gamma \delta$ as $\gamma$ is increased. Only when two eigenvalues collide (thereby creating a degeneracy), they might shoot off into the complex plane. This can be described as a spontaneous breaking of the $\mathbb{PT}$ symmetry and at these points the Liouvillian becomes non-diagonalizable \cite{Kanki2016}.

While the decay modes with eigenvalues on the real axis originate from purely diagonal operators, the perturbation does yield non-zero off-diagonal elements to first order in $\gamma$. Likewise, those on the vertical axis may have non-zero diagonal elements under the dissipative perturbation\footnote{Unless the unitary and dissipative parts of the Liouvillian commute with one another. In that case, diagonal and off-diagonal modes will remain separated. This would make the dynamics largely trivial, though.}. The claim that decoherence is purely determined by modes with decay rate $\gamma \delta$ is therefore only true for the asymptotic limit $\gamma \rightarrow 0$. But beyond this limit, the presence of weak symmetries can divide the spectrum in such a way that all decay modes on the real axis are confined to one symmetry sector.

One can understand this as follows: consider a weak symmetry $[\hat{\mathcal{L}}, \hat{O}] = 0$ where $\hat{O}$ has $n$ distinct eigenvalues, which label the different blocks that partition the space of operators. Unless the dissipation is fine-tuned in a very particular way, the Hamiltonian and dissipative parts of $\hat{\mathcal{L}}$ should separately commute with $\hat{O}$. This implies that $\hat{O}H = H$, such that the Hamiltonian is found in the sector corresponding to eigenvalue 1, i.e.\ the invariant subspace of $\hat{O}$. Writing $H = \sum_i \epsilon_i |\psi_i\ra \la \psi_i|$, the individual projectors onto the energy eigenstates must also be part of that sector, assuming that $\hat{O}$ is not specifically constructed to permute these different projectors (in which case it would be unlikely to commute with the dissipator). In other words, all operators that are diagonal in the energy eigenbasis are invariant under $\hat{O}$ and must belong to the same symmetry block. In the presence of $\mathbb{PT}$-symmetry, these are precisely the operators responsible for the eigenvalues on the real axis! As the dissipation is turned on perturbatively and the eigenvalues spread along the axis, these diagonal decay modes will be mixed with others (introducing off-diagonal components), but only those within the same symmetry sector. Due to the weak symmetry, the block structure is preserved.

We have shown that, for $\mathbb{PT}$-symmetric Liouvillian dynamics with a weak symmetry, all decay modes on the real axis belong to the same symmetry sector. As mentioned before, this becomes relevant when studying the time-evolution of the expectation value of observables. For observables outside out of this sector, with no components invariant under $\hat{O}$, the decay modes on the real axis do not contribute. In case of sufficiently weak system-bath coupling $\gamma$, all other decay modes lie on the vertical access and decay at the same rate. The result is an overall exponential damping factor, on top of the unitary dynamics of the closed system. The interplay between weak and $\mathbb{PT}$-symmetry, and its effect on the dynamics of observables, constitutes our main result. The rest of the paper is dedicated to a concrete example of the phenomenon.

\section{XXZ Heisenberg spin chain with bulk dephasing}\label{section:xxz}
As an example, we consider the spin-$\frac{1}{2}$ XXZ anisotropic Heisenberg chain, given by the Hamiltonian
\begin{align}
  H = J \sum_{i=1}^{N-1} \left(S_i^x S_{i+1}^x + S_i^y S_{i+1}^y + \Delta S_i^z S_{i+1}^z \right)
\end{align}
with coupling strength $J$, anisotropy $\Delta$, zero magnetic field and open boundary conditions. For the dissipative part, we consider bulk dephasing noise, defined by the $N$ jump operators $L_i = S_i^z$. This open quantum system cannot be solved by any known analytical methods, except in the $\Delta = 0$ limit where it can be mapped to a Hubbard model and is solvable by Bethe ansatz \cite{Essler_prosen}. Nonetheless there have been some good numerical studies on the system, in particular on the scaling of its dissipative gap \cite{Znidaric}.

Since the total magnetization $M = \sum_i S^z_i$ commutes with the Hamiltonian and with all jump operators $L_i$, it serves as the generator of a strong symmetry. This means that there are $2N-1$ magnetization blocks in $\mathcal{H}$ and the same number of diagonal blocks in $\mathcal{B(H)}$, each of which has its own steady state. These steady states are the maximally mixed states within each sector, as is easily checked by insertion into the Lindblad master equation. Thanks to the block structure, we can safely restrict ourselves to the zero-magnetization sector, which contains a lot of interesting physics. Note that the energy spectrum within this sector is non-degenerate, except at specific values of $\Delta$ corresponding to the XXZ model's roots of unity \cite{Deguchi2001}.

\begin{figure}[t!]
  \centering
    \includegraphics[width=\columnwidth]{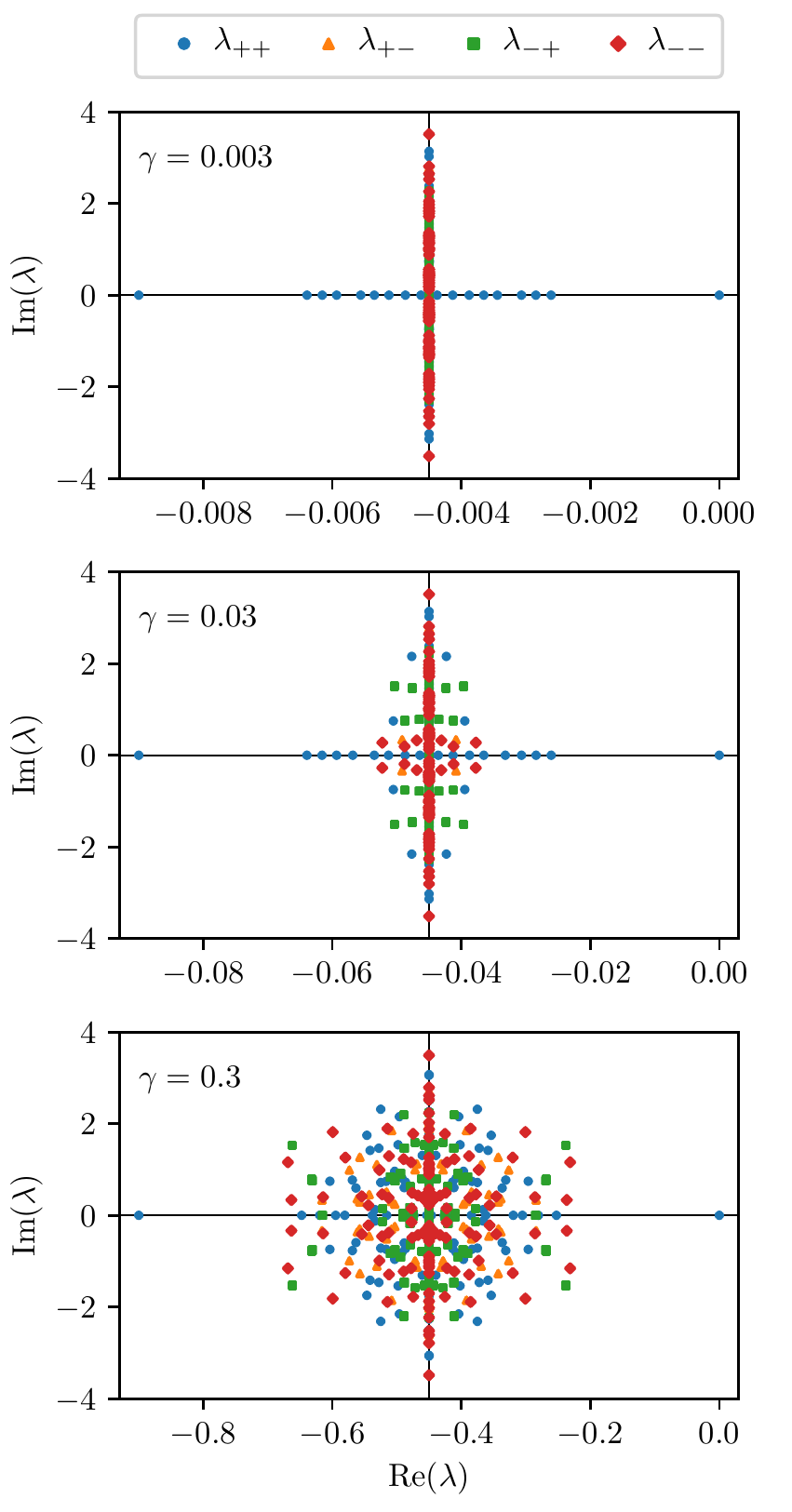}
    \caption{The spectrum of the Liouvillian for the dissipative XXZ Heisenberg chain with $N = 6$, $\Delta = 0.3$ and three different values of the system-bath coupling $\gamma$. The plot axes, as well as $\gamma$, are in units of the nearest-neighbor coupling $J$. Eigenvalues are labeled according to their symmetry sector $\hat{\mathcal{U}}_{p,q}$ with $p, q \in \{\pm 1\}$. At the top, $\gamma = 0.003 < \gamma_{PT} \approx 0.013$ shows all eigenvalues located on the two axes of reflection. Those on the real axis all belong to sector $\hat{\mathcal{U}}_{+,+}$. As $\gamma$ is increased, the $\mathbb{PT}$ symmetry is spontaneously broken and eigenvalues of all sectors move away from the vertical axes, into the complex plane.}
    \label{fig:spectrum}
\end{figure}

The zero-magnetization sector contains two additional weak symmetries, corresponding to spatial reflection and spin inversion.
\begin{alignat}{2}
  R &= \prod_{i=1}^{N/2}\Big(S^+_i S^-_{N+1-i} + &&S^-_i S^+_{N+1-i} + 2S^z_i S^z_{N+1-i} + \frac{1}{2}\id \Big) \nonumber \\
         &  &&\Rightarrow ~~~ R S^z_i R = S^z_{N+1-i} \\[6pt]
  F &= \prod_{i=1}^N \left(S^+_i + S^-_i\right) &&\Rightarrow ~~~ F S^z_i F = - S^z_i.
\end{alignat}
Both are parity operators, i.e.\ $R^2 = F^2 = \id$ with eigenvalues $\pm 1$. $R$ and $F$ commute with the Hamiltonian and which each other, but not with the individual jump operators. However, it is easy to check that the superoperators $\hat{R} \rho \equiv R \rho R$ and $\hat{F}\rho \equiv F \rho F$ do commute with the Liouvillian. Therefore the zero-magnetization sector of the Liouville space is split into four blocks $\hat{\mathcal{U}}_{p,q}$ labeled by the eigenvalues $p, q \in \{\pm 1\}$ of $\hat{R}$ and $\hat{F}$. The steady state, being proportional to the identity matrix, naturally is found in $\hat{\mathcal{U}}_{+,+}$. In fact, any decay mode that is purely diagonal in the energy eigenbasis will belong to this symmetry block. This can be seen as follows: since any energy eigenstate $|\psi\ra$ is also an eigenstate of $R$ and $F$ with eigenvalue $\pm 1$, we conclude that $|\psi\ra\la \psi|$ must be invariant under the superoperators $\hat{R}$ and $\hat{F}$. This is relevant, considering that the system is also $\mathbb{PT}$-symmetric.

The traceless part of the Liouvillian, as defined in (\ref{eq:tracelessL}), is given by
\begin{align}
  \hat{\mathcal{L}}'\rho = -i \left[H, \rho \right] + \gamma \sum_i S^z_i \rho S^z_i.
\end{align}
The parity superoperator $\hat{\mathcal{P}}$ is given by left-multiplication of the spin inversion $F$, such that $\hat{\mathcal{P}} \rho = F \rho$. It is now simple to check that the condition (\ref{eq:ptsymmetry}) for a $\mathbb{PT}$-symmetric Liouvillian is satisfied:
\begin{align}
  \begin{split}
  \mathcal{P}\mathcal{L'}\mathcal{P} \rho &= -i F \left[ H, F \rho \right] + \gamma \sum_i F S^z_i F \rho S^z_i\\  &= -i \left[ H, \rho \right] - \gamma \sum_i S^z_i \rho S^z_i = -(\mathcal{L}')^\+ \rho,
  \end{split}
\end{align}
Figure \ref{fig:spectrum} shows the Liouvillian spectrum for three values of $\gamma$. For sufficiently weak coupling, all eigenvalues are located along the two axes of reflection. Those along the real axis all correspond to decay modes in the $\hat{\mathcal{U}}_{+,+}$ symmetry block, which can be understood as follows: in the limit $\gamma \rightarrow 0$, these decay modes are purely diagonal in the energy eigenbasis and therefore even under $\hat{R}$ and $\hat{F}$. Because the dissipation preserves the symmetry structure in Liouville space, the modes must remain in the $\hat{\mathcal{U}}_{+,+}$ sector as the perturbation is turned on, even though they are no longer purely diagonal. In section \ref{section:neel}, we will see how this affects observables such as the staggered magnetization.

\begin{figure}[t]
  \centering
    \includegraphics[width=\columnwidth]{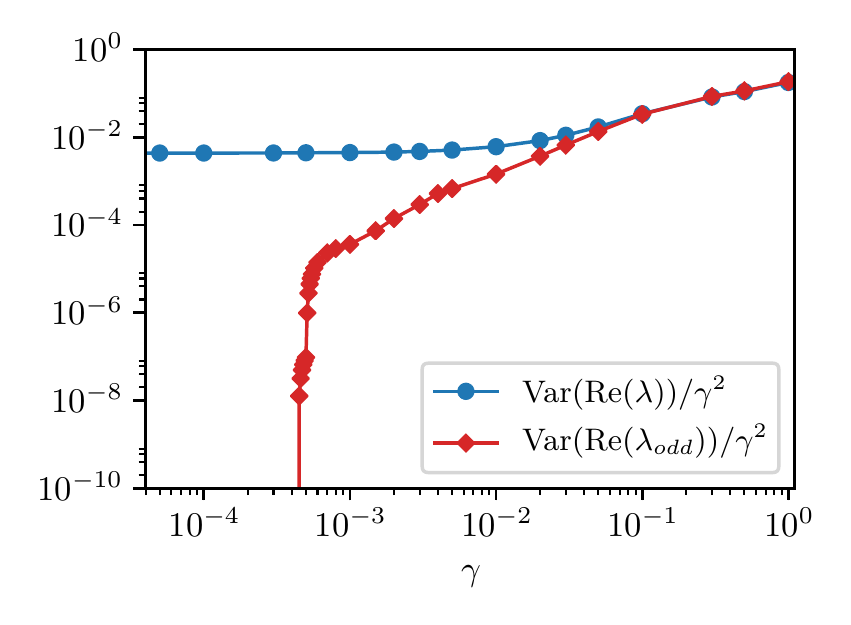}

    \caption{The variance of the real part of the Liouvillian spectrum as a function of the coupling strength $\gamma$, computed in units of $J$ for the dissipative XXZ Heisenberg chain with $N = 8$ and $\Delta = 0.3$. The (blue) circles indicate the variance over all eigenvalues, while the (red) diamonds include only those in the double-odd symmetry sector $\hat{\mathcal{U}}_{-,-}$. Values are rescaled by a factor $\gamma^{-2}$ to account for a uniform linear dependence on $\gamma$. A discontinuity around $\gamma_{PT} \approx 0.0003$ is clearly visible.}
    \label{fig:spectrum_variance}
\end{figure}

As $\gamma$ is further increased, the dynamics undergoes a transition where the $\mathbb{PT}$ symmetry is spontaneously broken and some of the eigenvalues leave the two axes. In Ref.\ [\onlinecite{Prosen2012}], an estimate is given for the critical coupling strength $\gamma_{PT}$ at which this happens. By computing the operator norm of the dissipator\footnote{As we are concerned with pure dephasing, the dissipator is diagonal in the local spin basis. This makes it straightforward to find the largest eigenvalue.} and estimating, in turn, the average density of states, we find the following expression for our model:
\begin{align}
  \gamma_{PT} \approx J \, \frac{(N-1)^2}{N}\, \binom{N}{N/2}^{-2}.
\end{align}
Unfortunately this quantity decays exponentially as $N$ becomes large. However, even for coupling strengths well above $\gamma_{PT}$, the effects of the $\mathbb{PT}$ symmetry remain visible. Figure \ref{fig:spectrum_variance} shows the spread in the real part of eigenvalues, both for all eigenvalues and for only those in the double-odd sector $\hat{\mathcal{U}}_{-,-}$. As can be seen, the variance within the odd sector is far below that of the total variance for a significant region of parameter space, even after the sharp jump at $\gamma = \gamma_{PT}$. This also ties into the results of Ref.\ [\onlinecite{Znidaric}], where a critical coupling $\gamma_c$ is described, at which the global dissipative gap switches from the even to the odd symmetry sector. This coupling $\gamma_c$ scales as $~ N^{-2}$, rather than exponentially.

\section{Staggered magnetization after a N\'eel quench}\label{section:neel}

The interplay of $\mathbb{PT}$ symmetry and weak parity symmetries results in an interesting structure within the Liouvillian spectrum of the XXZ chain with dephasing noise. To find out whether this is more than just a mathematical oddity, let us consider one of the natural observables for this system. The staggered magnetization is defined as
\begin{align}
  M_s = \frac{1}{N} \sum_{i=1}^N (-1)^i S_i^z.
\end{align}
Its expectation value is maximized in the N\'eel state, defined as $|\text{N\'eel}\ra = |\!\downarrow \uparrow \downarrow \uparrow\! \ldots \ra$ in the local spin basis. We can imagine preparing the system in the N\'eel state and looking at the evolution of the staggered magnetization after the state is released. This can be described as a quantum quench from the Ising antiferromagnet ($\Delta \rightarrow \infty$) to the XXZ model, which was studied numerically (in the absence of dissipation) in Refs.\ [\onlinecite{Barmettler,Barmettler2010}]. Since the N\'eel state has non-zero overlaps with all energy eigenstates in the zero-magnetization sector, the unitary dynamics at short times is extremely complex and impossible to study analytically, even using the tools of integrability. The numerics show that the staggered magnetization $M_s$ decays exponentially, modulated by an oscillation in the gapless regime. In the non-interacting limit ($\Delta = 0$), the decay becomes algebraic and is described exactly by a Bessel function.

Because the N\'eel quench provides such a rich unitary dynamics, it is well-suited to see the extreme effects of the symmetry structure within Liouville space. Naively, one would expect the dissipation to introduce many new timescales into the system, effectively destroying the characteristic behavior of the closed system. Looking back to Eq.\ (\ref{eq:obs_spectral}), the factor $\operatorname{tr}(v_m^\+ \rho(0))$ is non-zero for all decay modes, due to the nature of the N\'eel state. As it is an eigenstate of neither $R$ nor $F$, the density matrix $\rho(0)$ has components in all four symmetry blocks. The factor $\operatorname{tr}(O u_m)$, on the other hand, depends on the symmetry properties of the observable.

Assuming that $N$ is even, the staggered magnetization is antisymmetric under both of the parity symmetries:
\begin{align}
  R M_s R = F M_s F = -M_s \label{eq:MsSym}
\end{align}
and is therefore located within the $\hat{\mathcal{U}}_{-,-}$ symmetry block of Liouville space. It will be orthogonal under the Hilbert-Schmidt inner product to any decay modes within other sectors. As a result, only the decay modes in $\hat{\mathcal{U}}_{-,-}$ will contribute toward the time evolution of $\la M_s \ra$, regardless of the initial state. And thanks to the $\mathbb{PT}$ symmetry, for $\gamma < \gamma_{PT}$ all those modes have eigenvalues on the vertical symmetry axis and hence decay with the same rate $\delta$. The weak dephasing noise only introduces one new timescale after the N\'eel quench, yielding an overall exponential damping factor on top of the existing (unitary) exponential decay of the staggered magnetization.

This can be made more explicit by applying perturbation theory in $\gamma$ to Eq.\ (\ref{eq:obs_spectral}), expanding $\lambda_m$, $u_m$ and $v_m$. Since the perturbation does not mix modes from different symmetry sectors, the expansion only involves off-diagonal coherences and there are no degeneracies. The calculation is done in the appendix. In addition to the overall factor $e^{-\gamma \delta t}$, we see a $\gamma^2$ correction to the expectation value, due the shift of the decay modes along the vertical axis.

\begin{figure}[t]
  \centering
    \includegraphics[width=\columnwidth]{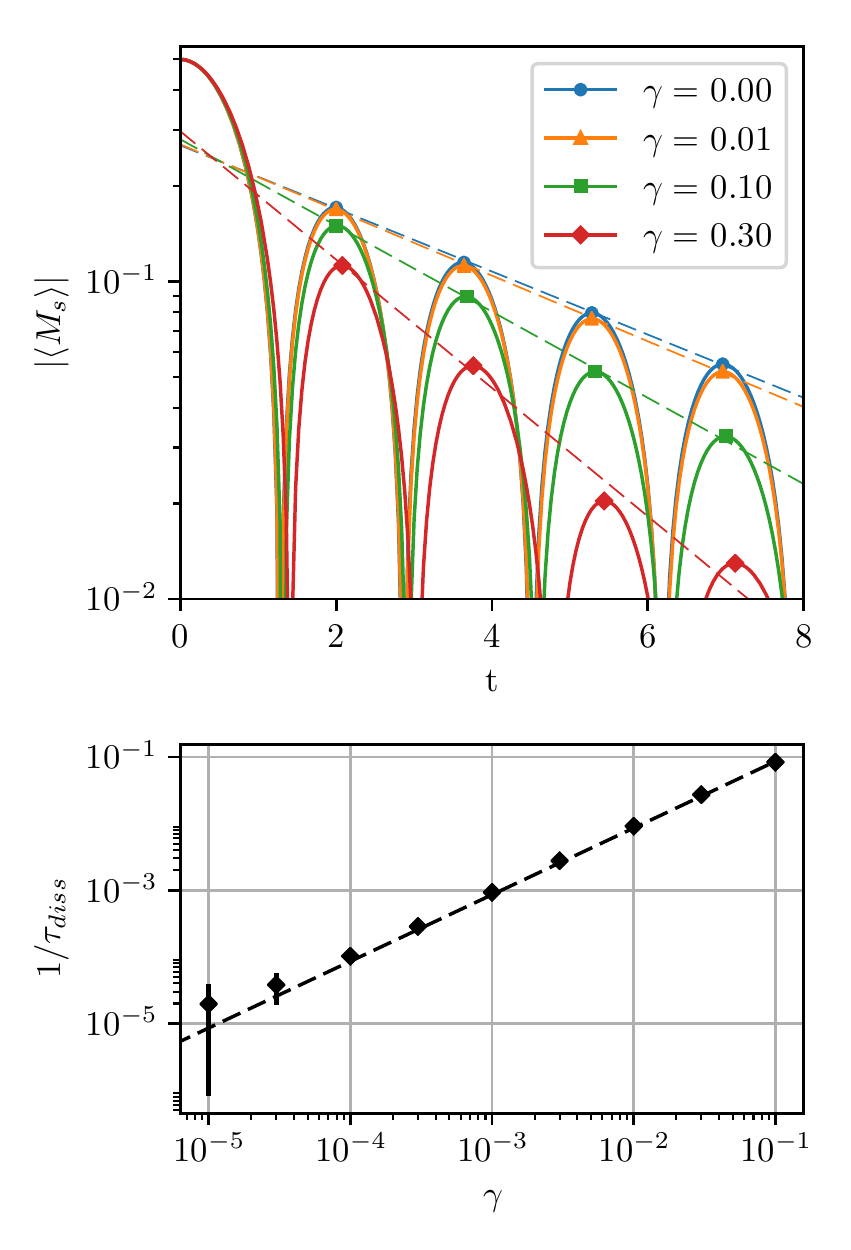}
\caption{The staggered magnetization after a N\'eel quench of the XXZ spin chain with $N = 12$ and $\Delta = 0.4$, for various values of the bulk dephasing strength $\gamma$ (in units of $J$). For the closed system and for weak dephasing, the expectation value shows an exponentially damped oscillation. The oscillatory behavior is largely unchanged for $\gamma < 0.1$. Dashed lines show exponential fits of the envelopes. The bottom panel shows the resulting decay rates due to the dephasing $1/\tau_{diss} = 1/\tau(\gamma) - 1/\tau(0)$, with error bars acquired from the exponential fit. It is clear that $1/\tau_{diss}$ is proportional to $\gamma$. Parameters for this computation were chosen to minimize finite-size effects.}
  \label{fig:ms_rates}
\end{figure}

We have numerically confirmed the above using a master equation solver \cite{qutip} within the relevant symmetry sector \cite{Sandvik} and the results are shown in figure \ref{fig:ms_rates}. Even for $\gamma$ much higher than $\gamma_{PT} \approx 10^{-5}$, the exponentially damped oscillations are preserved. The only effect of the bulk dephasing is an increase of the decay rate, proportional to $\gamma$, as predicted.

\section{Discussion}
We have shown that the effect of weak bulk dephasing on the staggered magnetization of the XXZ spin chain consists of a single exponential damping factor $e^{-\gamma \delta}$. This stems from the combination of various symmetries, acting on the different levels of a hierarchy of Hilbert spaces. On the level of quantum states, the conserved magnetization generates a strong symmetry, allowing a restriction to the zero-magnetization sector. On the level of operators, there are two weak symmetries in the form of reflection $\hat{R}$ and spin inversion $\hat{F}$, which divide the Liouvillian spectrum into four blocks $\hat{\mathcal{U}}_{p,q}$ with $p,q \in \{\pm 1\}$. And on the superoperator level, the $\mathbb{PT}$ symmetry of the Liouvillian forces its spectrum into a unique shape. The result is a spectral separation of the symmetry sectors, where all modes contributing to the staggered magnetization $M_s$ will decay at the same rate.

It is now interesting to define a general recipe, that can be applied to look for similar behavior in other systems. The required ingredients are a non-degenerate, $\mathbb{PT}$-symmetric Liouvillian and an observable of interest that is anti-symmetric under an additional weak parity symmetry. Since such anti-symmetries are built into the algebra of fermionic and spin systems, we suspect the phenomenon to be quite prevalent in such many-body models. Unfortunately it may be more difficult to find those properties in the simple bosonic systems that serve as popular models in quantum optics. Whether a $\mathbb{PT}$-symmetric Liouvillian is even possible in a purely bosonic system is an interesting open question.

In Ref.\ [\onlinecite{Prosen2012}], a boundary driven XXZ chain is given as an example of $\mathbb{PT}$ symmetry. There, one relevant observable is the spin current $J = i \sum_{i=1}^{N-1}\left(S_i^+ S_{i+1}^- - S_i^- S_{i+1}^+\right)$ which has vanishing diagonal elements in the energy eigenbasis, just like the staggered magnetization in our example above. The reason for this is that $J$ also is odd under the parity symmetries $R$ and $F$. However, this is not enough to ensure that the spin current relaxes with a uniform rate, except in the limit of $\gamma \rightarrow 0$. As we have seen, the decay modes on the real axis do have non-zero off-diagonal elements. Unlike the staggered magnetization under bulk dephasing, $J$ is not protected from these modes by a weak symmetry. That is because the driving of the spin chain is no longer symmetric under the superoperators $\hat{R}$ and $\hat{F}$ individually, but only under their product \cite{Buca2012}: $[\hat{\mathcal{L}}, \hat{R}\hat{F}] = 0$. The spin current is found in the even sector of this weak symmetry, and so are the decay modes on the real axis. It can easily be checked numerically that the contribution of these modes to the expectation value is small but non-zero. Particularly at long times, they may have a noticeable effect due to the slower decay rates. Another observable that \emph{is} confined to the odd symmetry sector is the total magnetization, which is not conserved by the boundary driving.

Also worth noting is that the addition of long-range interactions does not break any of the symmetries described for the XXZ spin chain. Going beyond nearest-neighbor coupling will affect $\gamma_{PT}$, but the structure of the symmetry sectors and the Liouvillian spectrum will remain the same. This is experimentally relevant in the context of trapped ions, which allow quantum simulation of spin chains with highly tunable long-range interactions \cite{Neyenhuis2017,Gras2014,Bermudez2017}. For such systems, bulk dephasing corresponds to local magnetic fluctuations within the trap, although there are also methods to control the dissipation \cite{Mueller2011}. It is our hope that the phenomenon of symmetry-protected coherent relaxation may be detectable in these kind of experiments.

\section{Acknowledgements}
This work is part of the Delta-ITP consortium, a program of the Netherlands Organization for Scientific Research (NWO) that is funded by the Dutch Ministry of Education, Culture and Science (OCW). The authors want to thank Enej Ilievski, Jean-S\'ebastien Caux and Dirk Schuricht for fruitful discussions and suggestions.

\appendix

\section{Perturbation theory of {$\la M_s \ra$}}
In this extra material, we will derive corrections to the staggered magnetization in the XXZ Heisenberg chain, resulting from weak bulk dephasing. We will draw heavily on the symmetry arguments from sections \ref{section:xxz} and \ref{section:neel}. As a starting point, consider Eq.\ (\ref{eq:obs_spectral}) in the $\gamma = 0$ case. Assuming non-degenerate energy eigenstate $H|\mu\ra = \epsilon_{\mu} |\mu \ra$, we find that $u_m = v_m = |\mu\ra \la \nu|$ with $\lambda_m = i(\epsilon_\mu - \epsilon_\nu)$. Since $M_s$ is confined to the double-odd symmetry sector $\hat{\mathcal{U}}_{-,-}$, only modes with $\mu \neq \nu$ need to be considered.

Now we can turn on the dissipation and apply perturbation theory to these off-diagonal modes. Writing the perturbation as
\begin{align}
  \hat{\mathcal{D}} \rho = \hat{\mathcal{D}}^\+ \rho = -\delta \rho + \sum_i S^z_i \rho S^z_i,
\end{align}
we find
\begin{align}
  \lambda_m &= \lambda_m^{(0)} + \gamma \operatorname{tr}(u_m^\+ \hat{\mathcal{D}} u_m) + \mathcal{O}(\gamma^2) \nonumber\\[5pt]
  &= i(\epsilon_\mu - \epsilon_\nu) - \gamma\delta + \gamma \sum_{i,j} \la \nu | S^z_i | \nu \ra \la \mu | S^z_j | \mu \ra + \mathcal{O}(\gamma^2) \nonumber\\
  &= i(\epsilon_\mu - \epsilon_\nu) - \gamma\delta + \mathcal{O}(\gamma^2),
\end{align}
where we have used that $\la \nu | S^z_i | \nu \ra = 0$. Similarly, the first-order correction to the decay modes becomes:
\begin{align}
  u_m \approx |\mu\ra \la \nu| - i \gamma \sum_{\substack{\mu', \nu' \\ \neq \mu, \nu}} \sum_{i,j} \frac{\la \mu'|S^z_i|\mu\ra \la \nu|S^z_j|\nu' \ra}{\epsilon_\mu - \epsilon_\nu - \epsilon_{\mu'} + \epsilon_{\nu'}} \, |\mu'\ra \la \nu'|. \nonumber
\end{align}
Note that the XXZ Hamiltonian is real and symmetric (most easily seen in the Pauli-representation of the local spin basis), which means that the matrix elements of $S^z_i$ are also real: $\la \mu | S^z_i | \nu \ra = \la \nu | S^z_i | \mu \ra$. Therefore, the first-order correction is purely imaginary.

The operators $M_s$ and $\rho_0 = |\text{N\'eel}\ra \la \text{N\'eel}|$ likewise have only real matrix elements.
Plugging the results above into Eq.\ (\ref{eq:obs_spectral}), we find:
\begin{align}
  &\la M_s(t) \ra = e^{-\gamma \delta t} \sum_{\mu, \nu \neq \mu} e^{it(\epsilon_\mu - \epsilon_\nu) + \mathcal{O}(\gamma^2)} \\
    \times &\Big(\la\mu|\rho_0|\nu\ra + i\gamma \sum_{\substack{\mu', \nu' \\ \neq \mu, \nu}} \sum_{i,j} \frac{\la \mu'|S^z_i|\mu\ra \la \nu|S^z_j|\nu' \ra}{\epsilon_\mu - \epsilon_\nu - \epsilon_{\mu'} + \epsilon_{\nu'}} \la \mu'|\rho_0|\nu'\ra \Big) \nonumber \\
  \times &\Big(\la\mu|M_s|\nu\ra - i\gamma \sum_{\substack{\mu', \nu' \\ \neq \mu, \nu}} \sum_{i,j} \frac{\la \mu'|S^z_i|\mu\ra \la \nu|S^z_j|\nu' \ra}{\epsilon_\mu - \epsilon_\nu - \epsilon_{\mu'} + \epsilon_{\nu'}} \la \mu'|M_s|\nu'\ra \Big) \nonumber
\end{align}
The cross terms, representing the corrections of order $\mathcal{O}(\gamma)$, are purely imaginary and cancel out when completing the sum over $\mu$ and $\nu$. As a result, the leading order correction due to the shifting decay modes is proportional to $\gamma^2$:
\begin{align}
  \la M_s(t) \ra = e^{-\gamma \delta t} \left(\la M_s(t) \ra_0 + \mathcal{O}(\gamma^2) \right),
\end{align}
where $\la M_s(t) \ra_0$ is the time-evolution for the closed system, as described in \cite{Barmettler}.

\bibliography{xxz_dephasing}

\end{document}